\documentstyle[prd,aps]{revtex}
\begin{document}
\iftrue 
\draft


\hfill {SMC-PHYS-154}

\hfill {hep-ph/9708242}

\centerline{\large\bf 	Complementarity between Gauge-Boson Compositeness}
\vskip 2pt
\centerline{\large\bf 		and Asymptotic Freedom}
\vskip 7pt
\centerline{		Keiichi Akama}
\centerline{\it	\small
		Department of Physics, Saitama Medical College,
		Kawakado, Moroyama, Saitama, 350-04, Japan}
\vskip 7pt
\centerline{		Takashi Hattori }
\centerline{\it	\small
		Department of Physics, Kanagawa Dental College,
		Inaoka-cho, Yokosuka, Kanagawa, 238, Japan}

\begin{abstract} 
\baselineskip 1pt
We derive and solve the compositeness condition 
	for the SU($N_c$) gauge boson
	at the next-to-leading order in $1/N_f$ 
	($N_f$ is the number of flavors)
	and the leading order in $\ln\Lambda ^2$
	($\Lambda $ is the compositeness scale)
	to obtain an expression for the gauge coupling constant 
	in terms of the compositeness scale.
It turns out that
	the argument of gauge-boson compositeness (with a large $\Lambda $)
	is successful only when $N_f/N_c>11/2$,
	in which the asymptotic freedom fails.
\end{abstract}


Recently, the possible compositeness of 
	gauge bosons \cite{cg,cgg,cgq,hidden,ind,emb}
	has attracted renewed attention 
	from both theoretical \cite{cgg2,cgq2,hid2,ind2}
	and phenomenological \cite{exp,pheno} points of view.
The experimental results at present do not exclude \cite{PDG}
	and may even suggest \cite{exp} the interesting possibility 
	that quarks, leptons and gauge bosons are composite \cite{pheno}.
Based on general theoretical analyses \cite{cgg,cgg2},
	this idea has been widely applied in various branches of physics.
In quark-lepton physics, various models have been considered 
	in terms of composite gauge bosons \cite{cgq,cgq2}.
In hadron physics, the vector mesons can be interpreted as gauge bosons 
	with hidden local symmetry \cite{hidden,hid2}.
The gauge fields induced in connection with the geometric phase
	(Berry phase) in molecular and other systems 
	are also expected to become dynamical 
	through the quantum effects of matter,
	and could be considered as composite gauge bosons \cite{ind,emb,ind2}.

A gauge boson interacting with matter
	can be interpreted as composite
	under the compositeness condition $Z_3=0$ \cite{cc},
	where $Z_3$ is the wave-function renormalization constant
	of the gauge boson.
Under this condition, the gauge field becomes an auxiliary field 
	without independent degrees of freedom \cite{GNKK}.
The quantum fluctuations, however, give rise to 
	a kinetic term of the gauge field, 
	so that a dynamical gauge boson is induced 
	as a composite of the matter fields.
The compositeness conditions which have been investigated so far 
	hold only in the large $N$ limit,
	where $N$ is the number of the matter fields
	coupling with the gauge boson.
However, in cases of practical interest, 
	$N$ is rather small.
Thus it is important to investigate the higher order effects in $1/N$.
In our previous papers, we derived the next-to-leading order (NLO)
	contributions to the compositeness conditions 
	in the Nambu-Jona-Lasinio model \cite{ccA}
	and the abelian gauge theory \cite{AH}
	at the leading order in $\ln(\Lambda ^2/m^2)$
	($\Lambda $ is the compositeness scale, and 
	$m$ is the constituent mass scale \cite{eps}).
In this paper, 
	we perform a similar investigation for the non-abelian gauge theory,
	and find that the argument of gauge-boson compositeness 
	(with a large $\Lambda $) is successful
	only when the number of flavors, $N_f$, is so large 
	that the gauge theory is not asymptotically free \cite{af}.
On the other hand, many people have argued that
	the asymptotically non-free theory 
	may encounter the problem of excessively large coupling constants
	at some ultraviolet energy scale,
	suggesting the necessity for some new physics 
	such as compositeness \cite{tc}.
Thus we observe a complementarity between 
	gauge-boson compositeness (with a large $\Lambda $), if any, 
	and asymptotic freedom in the gauge theories.

We consider the SU($N_c$) gauge theory 
	for the gauge boson $G_\mu ^a(a=1,\cdots ,N_c^2-1)$ 
	with $N_f$ fermionic matters $\psi _j(j=1,\cdots ,N_f)$: 
\begin{eqnarray} 
	{\cal L}^{(\Lambda )}&=& 
	-{1 \over 4} \left(  G_{\mu \nu }^a\right)  ^2  
	+ \sum _j \overline \psi _j 
	    \left(  i \not\!\partial - m 
		+ {1\over 2}g\lambda ^a \not\!\! G^a
	     \right)   \psi _j 
\cr &&
	-{1 \over 2\alpha } \left(  \partial ^\mu  G_\mu ^a\right)  ^2 
        +\partial ^\mu  \eta ^{a\dagger } 
         \left(  \partial _\mu  \eta ^a - g f^{abc} \eta ^b G_\mu ^c\right)  
\label{L}
\end{eqnarray} 
where	$G_{\mu \nu }^a = \partial _\mu G_\nu ^a - \partial _\nu  G_\mu ^a
                + g f^{abc} G_\mu ^b G_\nu ^c         $,
	$g$ is the coupling constant, $m$ is the mass of $\psi _j$,
	$\alpha $ is the gauge fixing parameter, and 
	$\eta ^a(a=1,\cdots ,N_c^2-1)$ is the Fadeev-Popov ghost.
The superscript $(\Lambda )$ of ${\cal L}$ indicates that
	the bare theory has a large but finite ultraviolet cutoff $\Lambda $.
In order to absorb the ultraviolet divergences 
	arising from quantum fluctuations,
	we renormalize the fields and coupling constant as \cite{Muta}
\begin{eqnarray} 
	{\cal L}^{(\Lambda )}&=& 
	-{1 \over 4} Z_3 \left(  G_{{\rm r}\mu \nu }^a\right)  ^2  
	+ \sum _j \overline \psi _{{\rm r}j} 
	   \left(  i Z_2 \not\!\partial - Z_m m_{\rm r} 
	      + {1\over 2}Z_1 g_{\rm r}\lambda ^a \not\!\! G_{\rm r}^a  
	    \right)  \psi _{{\rm r}j} 
\cr &&
	-{Z_3 \over 2Z_\alpha  \alpha _{\rm r}} 
	\left(  \partial ^\mu  G_{{\rm r}\mu }^a\right)  ^2 
        + Z_\eta  \partial ^\mu  \eta _{\rm r}^{a\dagger } 
         \left(  \partial _\mu  \eta _{\rm r}^a 
	 - {Z_1\over Z_2} g_{\rm r} f^{abc} \eta _{\rm r}^b G_{{\rm r}\mu }^c
	\right)  
\label{Lr}
\end{eqnarray} 
where	$G_{{\rm r}\mu \nu }^i 
	= \partial _\mu G_{{\rm r}\nu }^a - \partial _\nu  G_{{\rm r}\mu }^a,
	+ (Z_1/Z_2)g_{\rm r} f^{abc} G_{{\rm r}\mu }^b G_{{\rm r}\nu }^c $,
	the quantities with the index ``r" are the renormalized ones,
	and $Z_1$, $Z_2$, $Z_3$, $Z_\eta $, $Z_m$, and $Z_\alpha $
	are the renormalization constants.

Now let us impose the condition
\begin{eqnarray} 
	Z_3=0.
\label{cc}
\end{eqnarray} 
Under the condition (\ref{cc}), 
	Lagrangian (\ref{Lr}) involves no derivative of $G_{{\rm r}\mu }^i$ 
	so that the Euler equation with respect to $G_{{\rm r}\mu }^i$ 
	becomes a constraint, and 
	$G_{{\rm r}\mu }^i$ is an auxiliary field 
	without independent dynamical degrees of freedom.
To get the physical spectrum, however, we need to follow
	the usual renormalization procedure with (\ref{Lr}),
	where the physical gauge boson states do exist even 
	under the condition (\ref{cc}).
This means that the gauge boson is induced as a composite of matter fields,
	and hence (\ref{cc}) is called ``compositeness condition" \cite{cc}.

At the lowest order in $g_{\rm r}^2$, $Z_3$ is determined 
	so as to cancel the divergence in the one-loop diagrams 
	in Fig.\ \ref{f1} A--C. 
As is well known, this is given by \cite{Muta}
\begin{eqnarray} 
	Z_3=1-
	\left[ {2\over 3}N_f
	-\left(  {13\over 6}-{\alpha _{\rm r}\over 2}\right)  N_c\right] 
	g_{\rm r}^2I,
\label{Z0}
\end{eqnarray} 
where $I$ is the divergent integral, given 
	in the dimensional regularization by
	$I=1/16\pi ^2 \epsilon =1/8\pi ^2(4-d)$ 
	with the number of spacetime dimensions $d$.
A simple-minded application of the compositeness condition $Z_3=0$ gives
\begin{eqnarray} 
	g_{\rm r}^2=1\Bigg/
	\left[  {2\over 3}N_f
	-\left(  {13\over 6}-{\alpha _{\rm r}\over 2}\right)  N_c
	\right]  I . 
\label{g0}
\end{eqnarray} 
To avoid the absurdity of a vanishing coupling constant,
	we take the regularization 
	as an approximation to some physical cutoff,
	and fix $\epsilon $ at the non-vanishing value 
	$\epsilon =1/\ln(\Lambda ^2/m^2)$.
If we take the relation (\ref{g0}) as the leading one, 
	infinitely many diagrams of any higher order in $g_{\rm r}$
	have the same order of magnitude, and therefore 
	the usual perturbation expansion in $g_{\rm r}$ fails,
	as in the cases investigated in our previous works \cite{ccA,AH}.
Then we need to rely on a $1/N_f$ or $1/N_c$ expansion.
If $N_c$ is large,
	the solution for $g_{\rm r}^2$ in (\ref{g0}) becomes negative,
	and worse, 
	infinitely many higher loops with internal gauge-boson lines
	would also belong to the leading order in $1/N_c$.
Thus, the expansion in $1/N_c$ is not appropriate.
On the other hand, 
	$1/N_f$ can successfully classify diagrams, as we shall see below.
In this case, the leading order contribution comes 
	from the one-fermion-loop diagram (Fig.\ \ref{f1} A),
	and the one-boson-loop diagrams (Fig.\ \ref{f1} B,C) 
	belong to the next-to-leading order.
This harmonizes well with the physical picture 
	that the gauge boson is a composite of the fermions.

Now we turn to the NLO contributions in $1/N_f$.
In addition to the one-boson-loop diagrams B and C in Fig.\ \ref{f1}, 
	the multi-loop diagrams D--H in Fig.\ \ref{f1}
	belong to this order.
In Fig.\ \ref{f1}, the line of small circles  
	stands for the gauge boson propagator 
	with an arbitrary number of one-fermion-loops inserted.
The renormalization constant $Z_3$ should be chosen
	so as to cancel all the divergence in these diagrams.
Though the $n$-loop diagram diverges like $I^n$,
	it is suppressed by the factor $g^{2n}\propto 1/I^n$. 
Therefore, the leading divergent part of each diagram 
	is $O(I^n)O(I^{-n})\sim O(1)$, apart from the powers of $1/N_f$.
We denote this leading contribution of $O(1)$ from diagram 
	$X$(=D--H in Fig.\ \ref{f1}) to $Z_3$ by $Z(X,l)$,
	where $l$ is the number of the fermion loops per diagram.  
Different diagrams with the same $X$ and $l$ 
	give $Z(X,l)$ exactly the same contributions.
We call the number of such different diagrams 
	``multiplicity" of $X$ with $l$. 
The overlapping divergence in G and H is 
	separated into two parts:
``f" in respect to the fermion loop divergence
	at the three boson vertex part, and 
``m" in respect to the boson-fermion (mixed) loop divergence
	at the boson-fermion-fermion vertex part. 
We denote these contributions as 
$Z({\rm Gf},l)$,  
$Z({\rm Gm},l)$,  
$Z({\rm Hf},l)$, and  
$Z({\rm Hm},l)$.

The following properties are useful in the calculation.
(a)Due to the gauge symmetry,
	the leading divergences cancel each other in two diagrams:
	that with a fermion-loop inserted 
	at a three-gauge-boson vertex in some diagram
	and that with a fermion-loop inserted 
	into a gauge-boson propagator adjacent to the vertex. 
(b)Because the one-fermion-loop (denoted by $\Pi _0^{\mu \nu }$)
	inserted into a gauge-boson propagator (with momentum $q$)
	is divergenceless (i.e. $q_\mu \Pi _0^{\mu \nu }=0$),
	it diminishes the $\alpha _{\rm r}$-dependent part of the diagram.

The divergent contribution $Z(X,l)$ is separated into
 	the gauge independent part  $Z_0(X,l)$
	and the part $Z_\alpha (X,l)$ linear in $\alpha _{\rm r}$, 
	while the terms higher in $\alpha _{\rm r}$ are convergent.
We first consider the $\alpha _{\rm r}$-independent parts. 
Because the lowest-order fermion self-energy part and 
	boson-fermion-fermion vertex part converge in the Landau gauge,
	$Z_0({\rm D},l)$ and $Z_0({\rm E},l)$ have no leading divergence.
Because property (a) implies 
	$Z_0({\rm F},l)=-Z_0({\rm Gf},l)=Z_0({\rm Hf},l)$ for relevant $l$,
	and their multiplicities are $l+1$, $2l$, and $l-1$, respectively, 
	they cancel out when $l\geq 1$.
For $l=0$, diagrams G and H are absent,
	while diagram F is nothing but diagram B 
	(including the $\alpha _{\rm r}$-dependent part), 
	and, together with diagram C, contributes the second term
	in the square brackets in (\ref{Z0}).
Because property (a) implies $Z_0({\rm Gm},l)=-Z_0({\rm Hm},l)$ $(l\geq 2)$
	and their multiplicities are $2l$ and $2(l-1)$, respectively, 
	diagrams Gm and Hm contribute the following terms to $Z_3$:
\begin{eqnarray} 
	-\sum _{l=1}^\infty 
	{1\over l(l+1)}\left(  -{2\over 3}\right)  ^{l-1}
	N_c N_f^l (g_{\rm r}^2 I)^{l+1}.
\label{Za0}
\end{eqnarray} 
Then we consider the $\alpha _{\rm r}$-dependent parts. 
Due to property (b),
	diagrams D and E have contributions only when $l=1$,
	and F, G, and H have contributions
	only when no fermion-loop is inserted 
	on one of the two gauge-boson lines. 
The multiplicities of F, Gf, and Hf, are 1, 2, and 1, respectively,
	and those of Gm and Hm are 2 and 2, respectively, 
	for any relevant $l$.
Because property (a) implies 
	$Z_\alpha ({\rm F},l)=-Z_\alpha ({\rm Gf},l)=Z_\alpha ({\rm Hf},l)$
	and $Z_\alpha ({\rm Gm},l)=-Z_\alpha ({\rm Hm},l)$ for relevant $l$, 
	they cancel out for $l\geq 2$.
For $l=1$, diagram H is absent, and
	the contributions from diagrams D, E, F, Gf, and Gm with $l=1$ 
	to $Z_3$ are 
\begin{eqnarray} 
	-{1\over 3}\alpha _{\rm r}N_c N_f(g_{\rm r}^2I)^2 
\label{Za}
\end{eqnarray} 
	times $-1+1/N_c^2$, $-1/N_c^2$, $-1/2$, 1, and 3/2, 
	respectively, and hence (\ref{Za}) times 1 in total.
The contribution from $l=0$ has already been taken into account in (\ref{Z0}).
Next, we renormalize the subdiagram divergences by
	subtracting the divergent counter parts of 
	(i) each fermion loop inserted into the gauge boson lines in D--H, 
	(ii) the fermion self-energy part in D,
	(iii) the fermion-boson vertex part in E,
	(iv) the three-boson vertex part in Gf and Hf, and 
	(v)the boson-fermion-fermion vertex part in Gm and Hm.
The contributions from the counter parts cancel out for even $l$,
	and amount to minus twice the original terms for odd $l$.
Thus in total they contribute the following terms to $Z_3$: 
\begin{eqnarray} &&
	\sum _{l=1}^\infty 
	{1\over l(l+1)}\left(  {2\over 3}\right)  ^{l-1}
	N_c N_f^l (g_{\rm r}^2 I)^{l+1}
	+{1\over 3}\alpha _{\rm r}N_c N_f(g_{\rm r}^2I)^2. 
\label{Zs}
\end{eqnarray} 
Collecting all the contributions in 
	(\ref{Z0}) and (\ref{Zs}) together,
	we finally obtain the compact expression 
\begin{eqnarray} 
Z_3 &=& 1 - { 2\over 3} N_f g_{\rm r}^2 I
         + {11\over 3} N_c g_{\rm r}^2 I
         - {\alpha _{\rm r}\over 2} N_c g_{\rm r}^2 I 
	(1 - { 2\over 3} N_f g_{\rm r}^2 I)
\cr &&
   + {3\over 2} N_c ( {3\over 2N_f} - g_{\rm r}^2 I)
      \ln(1 - { 2\over 3} N_f g_{\rm r}^2 I)
    +O({1\over N_f^3}).
\label{Ztot}
\end{eqnarray}

The compositeness condition $Z_3=0$ with expression (\ref{Ztot})
	looks somewhat complex.
We can, however, solve it for $g_{\rm r}^2$
	by iterating the leading-order solution into itself.
The solution is rather simple:
\begin{eqnarray} 
	g_{\rm r}^2 = {3\over 2N_fI}
	\left[  1+{11N_c\over 2N_f}+O({1\over N_f^2})\right].
\label{g}
\end{eqnarray} 
The logarithmic term in (\ref{Ztot}) is suppressed in the solution (\ref{g})
	by the factor which vanishes with iteration of the leading solution.
It is interesting that the solution does not depend on the 
	gauge parameter $\alpha _{\rm r}$, 
	in spite of the fact that $Z_3$ does.
The $\alpha _{\rm r}$-dependent term in (\ref{Ztot}) 
	is also suppressed in the solution
	in the same way that the logarithmic term is.
The solution of the compositeness condition should be gauge-independent
	because it is a relation among physically observable quantities.
Note that we assumed a large but finite physical cutoff $\Lambda $, 
	by fixing $\epsilon $ at a non-vanishing value
	$\epsilon =1/\ln(\Lambda ^2/m^2)$.
Because the above argument relies on $1/N_f$ expansion including iteration,
	the absolute value of the next-to-leading contribution
	should not exceed that of the leading one.
If we apply it to (\ref{g}), we obtain
\begin{eqnarray} 
	N_f>{11N_c\over 2}.
\label{NfNc}
\end{eqnarray} 
The allowed region of $N_f/N_c$ by (\ref{NfNc})
	is complementary to that for asymptotic freedom 
	in the gauge theory \cite{af}.
When the gauge theory is asymptotically free,
	the next-to-leading contributions to the compositeness condition
	are so large that the gauge bosons cannot be composites 
	of the above type.
On the contrary, when the theory is asymptotically non-free, 
	the next-to-leading order contributions are suppressed,
	and the gauge bosons can be interpreted as composite \cite{largescale}.
As was shown in the previous paper \cite{AH},
	the NLO contribution is suppressed in the abelian gauge theory, 
	making the compositeness interpretation successful.
This is in accordance with the above-mentioned complementarity,
	since the abelian gauge theory is not asymptotically free.
In general, in the asymptotically non-free theories 
	with the beta function derived via perturbations in 
	the coupling constant,
	the running coupling constant diverges at some ultra-violet scale. 
Then we should introduce a momentum cutoff,
	since the theory gives nothing above that scale.
Some new physics such as compositeness is required to fill 
	in this blank \cite{tc}.

The marginal value $N_f/N_c=11/2$ for asymptotic freedom
	was derived via the renormalization group method 
	at the leading order in $g_{\rm r}^2$ (and exactly in $N_c$ and $N_f$).
All the one-loop diagrams and the renormalization constants
	$Z_1$, $Z_2$, and $Z_3$ were used there.
On the other hand, (\ref{NfNc}) 
	was derived via the NLO perturbation in $1/N_f$,
	and by using multi-loop boson-self-energy diagrams and $Z_3$ only.
The coincidence of the marginal values $N_f/N_c=11/2$
	may not be accidental, however, 
	because in the latter 
	the large-$\Lambda $ approximation implied small $g_{\rm r}$, and
	we implicitly used the other one-loop diagrams 
	(other than the boson self-energy) as subdiagrams, 
	and also used $Z_1$ and $Z_2$ (besides $Z_3$)
	to renormalize the divergences due to the subdiagrams,
	while in the former (the renormalization group method) 
	one-loop contributions at some scale
	could be interpreted as a sum of multi-loops at the other scale.
It would be interesting to confirm this complementarity at a higher order, 
	or in other models such as those with scalar matter,
	or to argue from a more general point of view covering all orders.
We expect that these methods and results concerning NLO compositeness
	and the concept of complementarity presented here 
	will be useful in pursuing 
	the composite dynamics of nuclei and hadrons, 
	and the possible compositeness of quarks, leptons, 
	gauge bosons and Higgs scalars, 
	as well as useful in induced gauge theories concerning molecular, 
	solid-state, and other systems.

We would like to thank Professor H. Terazawa for useful discussions.

\begin{figure}
\label{f1}
\end{figure}
\fi 
\Large
\unitlength=1.5pt

\def\hdots#1(#2,#3)#4{\multiput(#2,#3)(#1 4,0){#4}{\circle*{2}}}

\def\bigcirc (#1,#2,#3){\put(#1,#2){\circle{#3}}}
\def\fcircle (#1,#2,#3){\put(#1,#2){\circle*{#3}}}
\def\hcircles(#1,#2)#3{\multiput(#1,#2)(5,0){#3}{\circle{3}}}

\def\Floop#1{\hdots-(-17,0)4{\thicklines\bigcirc (0,0,28)}\hdots+(17,0)4
  \if1#1\bigcirc (-9, 6,3)\bigcirc (-5,3,3)\bigcirc (0,2,3)
        \bigcirc (5,3,3)\bigcirc (9, 6,3)\fi 
  \if2#1\bigcirc (0,-10,3)\bigcirc (0,-5,3)\bigcirc (0,0,3)
        \bigcirc (0,5,3)\bigcirc (0,10,3)\fi 
}
\def\Gloop{ \hdots-(-17,0)4\hdots+(17,0)4
  \fcircle (-12.4, 4,2)\fcircle (-10.5, 7.6,2)\fcircle (-7.6, 10.5,2)
  \fcircle (-4, 12.4,2)\fcircle (0, 13,2)
  \fcircle (-12.4,-4,2)\fcircle (-10.5,-7.6,2)
  \fcircle (-7.6,-10.5,2)\fcircle (-4,-12.4,2)\fcircle (0,-13,2)
  \fcircle ( 12.4, 4,2)\fcircle ( 10.5, 7.6,2)
  \fcircle ( 7.6, 10.5,2)\fcircle ( 4, 12.4,2)\fcircle ( 13,0,2)
  \fcircle ( 12.4,-4,2)\fcircle ( 10.5,-7.6,2)
  \fcircle ( 7.6,-10.5,2)\fcircle ( 4,-12.4,2)\fcircle (-13,0,2)
}
\def\FPloop{ \hdots-(-17,0)4\hdots+(17,0)4
  \fcircle (-12, 5,1)\fcircle (-9.2, 9.2,1)
  \fcircle (-5, 12,1)\fcircle (0, 13,1)
  \fcircle (-12,-5,1)\fcircle (-9.2,-9.2,1)
  \fcircle (-5,-12,1)\fcircle (0,-13,1)
  \fcircle ( 12, 5,1)\fcircle ( 9.2, 9.2,1)
  \fcircle ( 5, 12,1)\fcircle ( 13,0,1)
  \fcircle ( 12,-5,1)\fcircle ( 9.2,-9.2,1)
  \fcircle ( 5,-12,1)\fcircle (-13,0,1)  
}
\def\Bloop#1#2{ 
  \bigcirc (-10, 8.4,3)\bigcirc (-5.5, 11.8,3)
  \bigcirc (0, 13,3)\bigcirc ( 5.5, 11.8,3)\bigcirc ( 10, 8.4,3)
  \bigcirc (-10,-8.4,3)\bigcirc (-5.5,-11.8,3)
  \bigcirc (0,-13,3)\bigcirc ( 5.5,-11.8,3)\bigcirc ( 10,-8.4,3)
  \if1#1{\thicklines\bigcirc (-13,0,12)}\hdots-(-21,0)3 
    \else \bigcirc (-12.5,-3.4,3)\bigcirc (-12.5,3.4,3)\hdots-(-13,0)5\fi 
  \if1#2{\thicklines\bigcirc ( 13,0,12)}\hdots+(+21,0)3 
    \else \bigcirc (12.5,-3.4,3)\bigcirc (12.5,3.4,3)\hdots+(+13,0)5\fi 
}
\def\F#1{\put(-30,7){#1}}
\begin{center}
\begin{picture}(10,10)
\put(-70,- 40){\F{A}\Floop0 }
\put(  0,- 40){\F{B}\Gloop  }
\put( 70,- 40){\F{C}\FPloop }
\put(-70,- 80){\F{D}\Floop1 }
\put(  0,- 80){\F{E}\Floop2 }
\put(-70,-120){\F{F}\Bloop00}
\put(  0,-120){\F{G}\Bloop01}
\put( 70,-120){\F{H}\Bloop11}
\put(-12,-180){FIG. 1} 
\end{picture}
\end{center}

\vskip 10cm
\leftskip3cm\rightskip3.5cm
\noindent
\small
The gauge-boson self-energy parts at the leading order (A)
and at the next-to-leading order (B--H) in $1/N_f$.
The solid, dotted, and small-dotted lines 
	indicate the fermion, gauge-boson, and Fadeev-Popov ghost 
	propagators, respectively.
The line of small circles 
	stands for the gauge boson propagator 
	with an arbitrary number of fermion-loops inserted.


\begin{thebibliography}{99}
\bibitem{eps}
In practice, we use the dimensional regularization 
	with correspondence $\epsilon =(4-d)/2=1/\ln(\Lambda ^2/m^2)$,
	where $d$ is the number of spacetime dimensions.
\bibitem{largescale}
This does not eliminate the possibility that the composite gauge boson
	with a low $\Lambda $ is compatible with asymptotic freedom.
\end{thebibliography}
\end{document}